\documentclass[superscriptaddress,twocolumn,showpacs,prl,aps,floatfix]{revtex4}
\usepackage{graphicx}
\usepackage{color,transparent}
\usepackage{amssymb,amsmath}
\usepackage{array}
\usepackage{dcolumn}
\usepackage{xcolor}
\usepackage[normalem]{ulem}
\newcommand\st{\bgroup\markoverwith
  {\textcolor{blue}{\rule[.35ex]{5pt}{1.1pt}}}\ULon}

\begin{document}

\title{Can we prevent the ``dead layer'' formation at manganite interfaces?}

\author{Ay\c seg\"ul Beg\"um Ko\c cak}
\affiliation{Institut N\'eel, CNRS UPR 2940,  Grenoble, France} 
\affiliation{Institut National Polytechnique de Grenoble, France}
\affiliation{Theoretical Materials Physics, Q-Mat, CESAM, Universit\'e
  de Li\`ege, 4000, Sart Tilman, Belgium}
\affiliation{Institut Laue Langevin, Grenoble, France}

\author{Julien Varignon} 
\affiliation{Theoretical Materials Physics, Q-Mat, CESAM, Universit\'e
  de Li\`ege, 4000, Sart Tilman, Belgium}

\author{S\'ebastien Lemal} 
\affiliation{Theoretical Materials Physics, Q-Mat, CESAM, Universit\'e
  de Li\`ege, 4000, Sart Tilman, Belgium}

\author{Philippe Ghosez} 
\affiliation{Theoretical Materials Physics, Q-Mat, CESAM, Universit\'e
  de Li\`ege, 4000, Sart Tilman, Belgium}

 \author{Marie-Bernadette Lepetit}
\affiliation{Institut N\'eel, CNRS UPR 2940,  Grenoble, France} 
\affiliation{Institut Laue Langevin, Grenoble, France}


\begin{abstract}
  The present work theoretically studies the possibility to hinder the
  formation of a ``dead'' layer at the interfaces in manganite
  superlattices. We showed that this goal can be reached by using
  alkaline-earth simple oxides as alternating layers in very thin
  superlattices. Indeed, such alternating layer promotes the
  contraction of manganite layers at the interfaces 
  and $d_{x^2-y^2}$ preferred $e_g$ orbital occupancy, while
  Boltzman's transport calculations show an increase in
  conductivity. This result hold for different manganites, different
  alkaline-earth simple oxides as well as different thicknesses of the
  two layers.
\end{abstract}

\maketitle
Interfaces between perovskite oxides have been  the subject of intense research over the last decade. The first reason is the
outstanding properties discovered in such interfaces~; let us only
cite the superconductivity discovered at the interface between two
band insulators such as $\rm SrTiO_3$ (STO) and $\rm
LaAlO_3$~\cite{supra}. Another reason is
the potential applications of such properties.  Manganite-based
devices using tunnel junctions are actively studied for the design of
spin valves or spin injectors. Such junctions present a high degree of
spin polarization and robust magnetic properties at the interface
between the manganite and the barrier.  The main problem that has hindered the development of the such devices is
the formation of a so-called "dead-layer"~\cite{DLYAMAD,DLRAME},
below a critical distance to the interface. In such layers the
 magneto-transport properties are strongly depressed. The present paper
proposes a possible solution to this critical problem with not
only a set of criteria to design appropriate barriers, but also a
detailed study of a realistic example.

Manganites are known to be ferromagnetic metals over a large range of
their phase diagram, and to present colossal magneto-resistance
effects. Indeed, the record value is of over 14 orders of magnitude in
resistivity change under magnetic field~\cite{MRrecord}).  Their
transport and magnetic properties are controlled by small atomic
displacements, allowing potential pathways to tune their properties
using interfaces in very thin films and hetero-structures (see
for instance Ref.~\onlinecite{TER,BT,accu,strainmagn}, and
Ref.~\onlinecite{revuebeigi} for a recent review).
Unfortunately, a loss of magnetization and metalicity, also called
``dead layer'', is observed over a thickness of few unit cells (u.c.) at the
interface of ferromagnetic manganites, such as $\rm La_{2/3}
Sr_{1/3}MnO_3$ (LSMO) or $ \rm La_{2/3}Ca_{1/3}MnO_3$ (LCMO), and most
perovskite substrates or combined layers~\cite{DLYAMAD,DLRAME}.
This ``dead layer'' phenomenon has been the subject of many
  interpretations such as (i) homogeneous substrate
  strain~\cite{Ziese}, (ii) electronic and/or chemical phase
  separation~\cite{Fontcuberta} related to structural inhomogeneities
  at the interface~\cite{BIBES} or uncontrolled
  stoichiometry~\cite{Kourkoutis}, (iii) manganese $e_g$ orbital
  reconstruction that may induce C-type
  antiferromagnetism~\cite{TEBANO_LD,TEBANO_ARPES} and can be
  attributed to a weak delocalization at the interfaces~\cite{LSMO}.
  The first hypothesis is incoherent with the relaxation rate of the
  substrate strain, shown to be larger than
  1000\,\AA{}~\cite{Fontcuberta}, while a drastic change in the
  transport properties~\cite{DLRAME,TEBANO_LD,Kourkoutis} is observed
  at the STO interface, for films thinner then a few u.c.
  In this work we would like to work with perfect interfaces, we will
  thus not consider the consequences of inperfectly grown interfaces
  and  focus on the last hypothesis of enhanced $d_{z^2}$
  occupancy at the interface. Such a behavior was attributed to an
  energetic lowering of the $d_{z^2}$ orbital over the $d_{x^2-y^2}$
  at the interface, due to a weak delocalization of the former through
  the interface~\cite{LSMO}.

Ferromagnetic manganites (and related hetero-structures) of general
formula $\rm La_{1-x} A_{x}MnO_3$ ($\rm A$ a divalent cation, $x$ in
the approximate range $0.2<x<0.45$) crystallize in a $\rm ABO_3$ perovskite
structure~\cite{struct_LSMO,struct_LCMO}, with the Mn occupying the B
site. The Mn atom is thus in an octahedron environment, which induces
an energetic splitting of the Mn $3d$ orbitals into a $t_{2g}$
---~threefold degenerated~--- low energy set, and a $e_g$ ---~twofold
degenerated~--- high energy one (ideal case). Moreover, the Mn atoms
are in a mixed valence ionic state ($\rm Mn^{3+x}$), with a $3d^{4-x}$
high spin orbital occupancy. As a result, the two $e_g$ orbitals,
$d_{z^2}$ and $d_{x^2-y^2}$, are partially occupied by $1-x$ electron
which may delocalize (the $d_{z^2}$ electrons along the {\bf c}
direction and the $d_{x^2-y^2}$ ones in the ({\bf a},{\bf b})
plane). This delocalization is energetically very favorable, but will
only occur when the spins of neighboring Mn ions are ferromagnetically
aligned. In such a case the delocalization energy gain overcomes the
antiferromagnetic exchange interaction between localized ions, and
imposes a ferromagnetic ordering (double exchange
mechanism)~\cite{dble-xchg}.

In bulk materials the $e_g$ electrons are shared between the two
orbitals with equivalent occupancies. It results in magnetic ordering
and delocalization occurring in all directions.  In very thin films,
however, the out-of-plane direction, {\bf c}, spans only over a few
u.c., and thus the thermodynamic limit is only obtained in the
in-plane, ({\bf a},{\bf b}), directions. It is thus of crucial
importance for the magnetic and transport properties to maximize the
$d_{x^2-y^2}$ orbital occupancy, responsible for the in-plane
delocalization and thus magnetic and transport properties. 

When an interface ``dead layer'' is present, the $d_{z^2}$ electrons
of the manganite delocalize in the empty $d_{z^2}$ orbitals of the
substrate or of the alternating layer (typically $\rm SrTiO_3$, $\rm
BaTiO_3$ or similar compounds). Even if weak (about one or two tenth
of an electron~\cite{LSMO}), at the interfaces this delocalization
energetically favors the $d_{z^2}$ orbital occupancy over the
$d_{x^2-y^2}$ one. It results in a Jahn-Teller distortion of the $\rm
MnO_6$ octahedron, with a splitting of the $e_g$
degeneracy~\cite{BTO_LSMO} ($\varepsilon_{z^2} <
\varepsilon_{x^2-y^2}$). The in-plane delocalization is thus hindered
(at least by carrier density reduction).  Consequently, the
characteristics of the ``dead layer'' phenomenon appear, (reduced
ferromagnetic spin arrangement and
conductivity~\cite{TEBANO_RM,DLRAME}).  Such a Jahn-Teller
distortion induces a small increase of the {\bf c}
parameter~\cite{HERGER,LSMO}, that can be fully attributed to
the delocalization mechanism at the interfaces,  as
strain effects tend to reduce {\bf c}. Indeed, on a STO substrate, manganites such as LSMO or LCMO are under tensile
strain ($\rm {\bf a}_{\rm STO}=3.905\rm \AA{}$~\cite{Struct_STO},
  ${\bf a}_{\rm LSMO}=3.880\rm\AA$~\cite{struct_LSMO}, ${\bf a}_{\rm
    LCMO}=3.86\rm\AA$~\cite{struct_LCMO} yielding a $1\%$ strain on
  LSMO and $2\%$ on LCMO), known to favor a reduction of {\bf c}.
In order to prevent the formation of a ``dead layer'', one thus needs
to interface the manganite with an alternating layer material
hindering the delocalization between the different layers.

The purpose of this work is thus to investigate, using first-principle
calculations, possible candidates for such alternating layers. We will
focus on LSMO-based hetero-structures over a $\rm SrTiO_3$ substrate.

The first idea that may come to mind, is to use alternating layers
with totally-filled $d$-shells, and a tetragonal
structure. Indeed, the latter was shown to be crucial
in order to prevent the rhombohedral distortion in the manganite
layer, but rather favor a tetragonal
one~\cite{BTO_LSMO}. Let us remember that a tetragonal distortion, associated with a {\bf c} parameter
contraction, allows an enhancement of the $d_{x^2-y^2}$ orbital
occupancy and thus of the desired properties.  One could therefore
think of materials such as the $\rm BaSnO_3$ compound. Unfortunately,
our test calculations on such hetero-structures exhibited a weak
electron delocalization, from the Sn filled $d_{z^2}$ orbitals towards
the Mn partially occupied ones, very similar to what we observed
  in our calculations on BTO/LSMO~\cite{BTO_LSMO} or STO/LSMO
  heterostructures (that exhibit a JT distortion of $\sim 1.04$ in the
  interface layer, a dominant $d_{z^2}$ occupancy and a weak $d_{z^2}$
  delocalization in the Ti orbitals.).  This delocalization is
associated with an increase of the Mn $d_{z^2}$ orbital occupancy, and
a Jahn-Teller distortion. One can thus expect such hetero-structures
to exhibit a ``dead layer'' phenomenon.

\begin{figure}[h]
\resizebox{8cm}{!}{\includegraphics{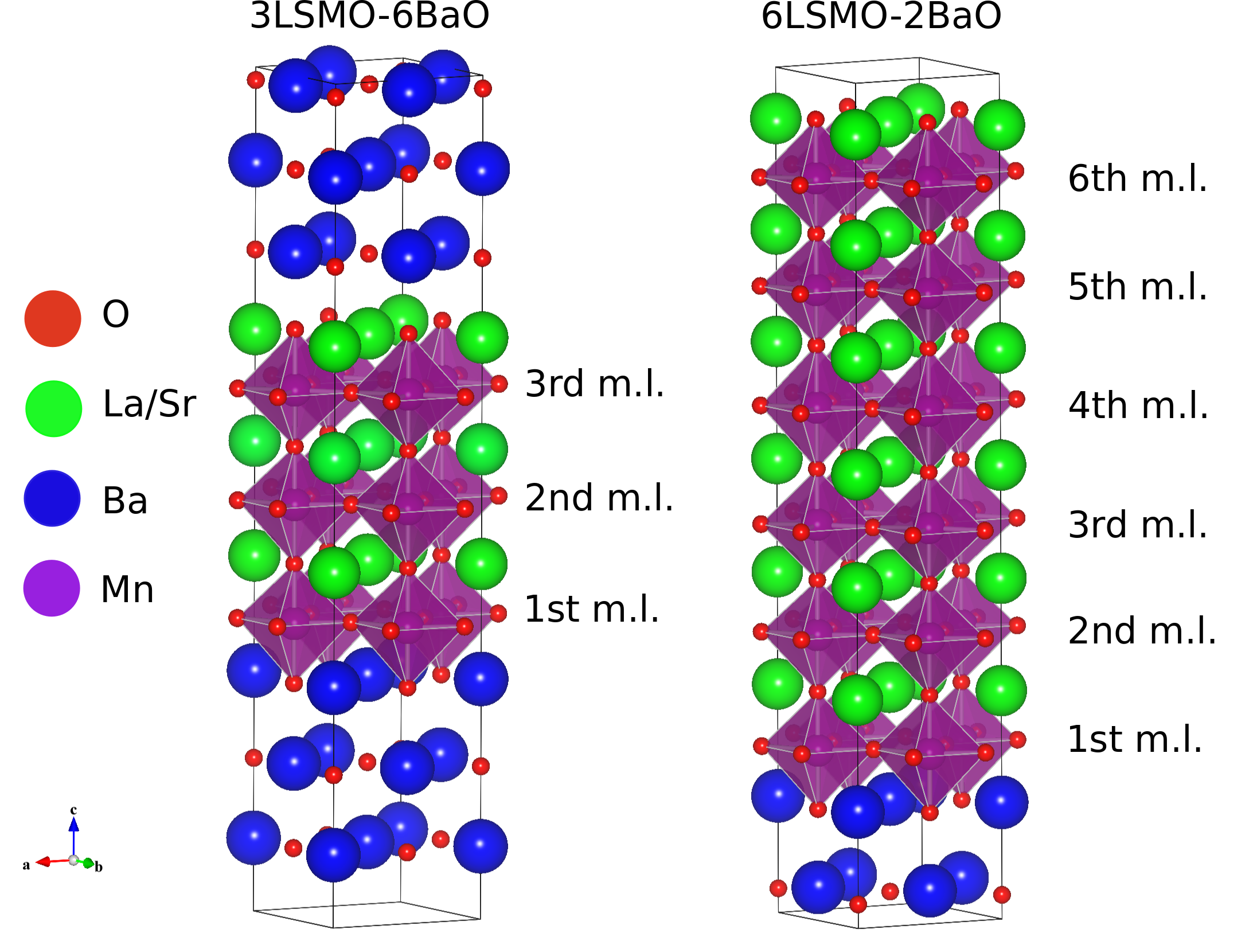}}
  \caption{(color online) Schematic representation of the $\rm
    [La_{2/3}Sr_{1/3}MnO_3]_3[AO]_6$ and
    $\rm [La_{2/3}Sr_{1/3}MnO_3]_6[AO]_2$ superlattices.}
    \label{fig:SR}
\end{figure}

Another way to prevent the inter-layers delocalization of the LSMO Mn
$d_{z^2}$ orbitals is to totally avoid $d$ orbitals, in the
alternating layer material. The requirement for the alternating layer
should thus be i) no $d$ orbitals, ii) a tetragonal or cubic structure,
and iii) a compound allowing perfect epitaxy with the manganite layer.
Fulfilling all those requirements are the simple alkaline-earth
oxides, and more specifically the $\rm BaO$ compound. Indeed, the
mismatch between BaO and LSMO is only of 0.7\%, and between BaO and
the STO substrate 0.3\%.  Of course the epitaxy imposes a $\rm BaO$
unit cell ($Fm\bar 3m$ cubic group~\cite{BaO}) rotated in-plane by an
angle of $45^\circ$~\cite{BaOrot}, compared to the manganite unit cell (see
figure~\ref{fig:SR}).

We thus studied, using first-principles calculations, $\rm [La_{2/3}
Sr_{1/3}MnO_3]_n[BaO]_p$ superlattices on a STO substrate,
alternating a few u.c. of manganite and of simple Barium
oxide. Superlattices with other alkaline-earth oxides were also
investigated to see whether the results are resilient to a change in
the alternating layer, despite their unrealistic strain
values~\cite{note}.

We performed geometry optimizations for the different superlattices,
using periodic density functional calculations. Since epitaxial films
normally follow the structure of the substrate, we imposed to our
optimizations to keep the substrate in-plane lattice constants
(optimized using the same computational parameters).  The
alkaline-earth oxides are strong insulators, while the manganite
layers are expected to be metallic, one thus needs to choose a
functional that properly positions the metal Fermi level with respect
to the insulator gap. We  used the B1WC hybrid
functional~\cite{B1WC}, that was specifically designed to properly treat
both gaps and weak distortions, two key components in the present
systems.  The calculations were done using the CRYSTAL
package~\cite{Crystal}, with the basis sets and effective core
pseudopotentials (ECP) of ref.~\onlinecite{Bases}.  As the LSMO A-site
cations disorder is difficult to treat within periodic calculations, we
run a set of calculations with different orderings, using true atoms
or average ones.  The average ions were modeled as in
reference~\onlinecite{BTO_LSMO}, that is using ordered cations ECPs
but with averaged effective nuclear charges. The effect of these
average charges is to hinder possible electronic localization induced
by the cation orders. Unless specified, we will only present results
that are independent of the cation order or model. Finally we used a
$\sqrt{2} a\times \sqrt{2}a\times c$ unit cell in order to allow
octahedra rotations and in-plane antiferromagnetic (AFM) ordering (see
figure~\ref{fig:SR}).


We first studied the $\rm [La_{2/3}Sr_{1/3}MnO_3]_3[BaO]_6$
superlattice, using all 6 cations models and different magnetic
orders~; that is ferromagnetic (FM), A-type AFM (in-plane FM and
out-of-plane AFM), C-type AFM (in-plane AFM and out-of-plane FM) and
G-type AFM (in-plane and out-of-plane AFM).

Notice that in what follows we consider superlattices with
stoichiometric layers yielding, in most cases, asymmetric interfaces
as it would be in real heterostructures. As it will be discussed in
the last Section, this has however no direct impact on our
conclusions.

Our calculations showed that the magnetic ground state always imposes
a FM in-plane order and a total net magnetic moment. The two
out-of-plane magnetic arrangements are quasi-degenerate within DFT
error bars. Indeed, the energy difference per LSMO u.c. (or
equivalently per Mn), between FM and A-type AFM orders, is in average
8\,meV/u.c., with a mean deviation of 16\,meV. This is smaller than the
room temperature ($k_BT\sim 25$\,meV). Whether the DFT ground state is
the FM or the A-type AFM configuration depends on the specific cation
ordering.  The in-plane AFM ordered states are much higher in energy,
ranging between 130\,meV and 210\,meV above the ground states.

\begin{figure}[h!]
\resizebox{7cm}{!}{\includegraphics{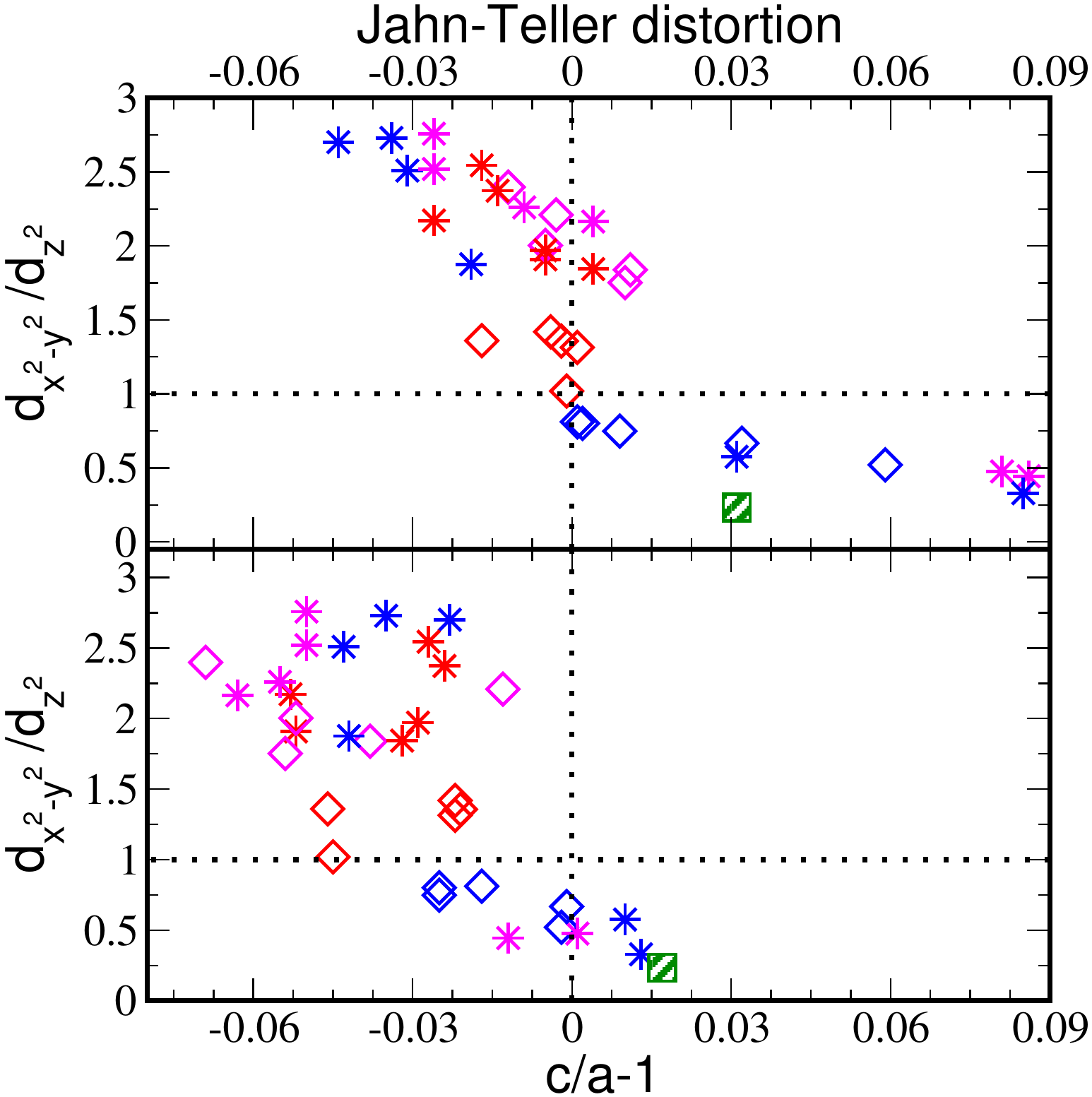}}
\caption{(color online) $\eta(d_{x^2-y^2})/\eta(d_{z^2})$ ratio of the
  $d_{x^2-y^2}$ and $d_{z^2}$ orbitals M\"ulliken occupancies in each
  mono-layer as a function of the Jahn-Teller distortion (JTd)
  (measured as $d_{\rm OO}/a-1$, with $d_{\rm OO}$ the out-of-plane
  O-(Mn)-O distance) and of $c/a$ ($c$ being measured as the
  perovskite A-sites distance).  Red and pink symbols refer to the
  interfacial mono-layers, blue symbols to the central ones. Diamonds
  are for the FM order and stars for the A-type AFM one. The
    green dashed squares show the experimental values of LSMO over STO
    for 6 m.l. thin films exhibiting a dead layer.  The Jahn-Teller
    distortion and $c/a$ ratios are extracted from the cumulative
    displacements in Ref.~\onlinecite{HERGER} and the
    $\eta(d_{x^2-y^2})/\eta(d_{z^2})$ ratio is extracted from linear
    dichroism experiments of Ref.~\onlinecite{TEBANO_LD}. Bulk LSMO
    corresponds to the cross point between the dashed lines.}
  \label{f:pi_ca}
\end{figure}
Figure~\ref{f:pi_ca} pictures the statistics of both the Jahn-Teller
distortion (JTd) (measured as $d_{\rm OO}/a-1$, with $d_{\rm OO}$ the
out-of-plane O-(Mn)-O distance) and $c/a$ ($c$ being the perovskite
A-sites distances) as a function of the $e_g$ orbitals-occupancies
ratio.  Results are given for each mono-layer and the two possible
ground states. Before analyzing the results let us keep in mind that
the two interface layers are non equivalent, since one corresponds to
a (La/Sr,O)--(BaO) interface and the other to a (${\rm MnO_2}$)--(BaO)
  interface.

  One sees immediately that all three layers are compressed along the
  {\bf c} direction, except for the central layer in two AFM
  calculations.  Similarly, the Mn-octahedra of the interface
  mono-layers are compressed along  {\bf c} and
  display a dominant $d_{x^2-y^2}$ orbital occupancy, favorable
    to the searched magnetic semi-metal behavior. Only
    the central monolayer exhibits sometimes an elongation of the
    Mn-octahedra, with a dominant $d_{z^2}$ occupancy. This behavior is the
    exact opposite of what is found in $\rm [LSMO]_3[BTO]_3$
    superlattices~\cite{BTO_LSMO,beigi2}, in which the interface layers are
    elongated with a dominant $d_{z^2}$ occupancy, responsible for the
    “dead” layer behavior. One should also point out that octahedra
  rotations are essentially negligible in these
  hetero-structures (rotation values less than $1^\circ$).

\begin{figure}[h]
\resizebox{8cm}{!}{\includegraphics{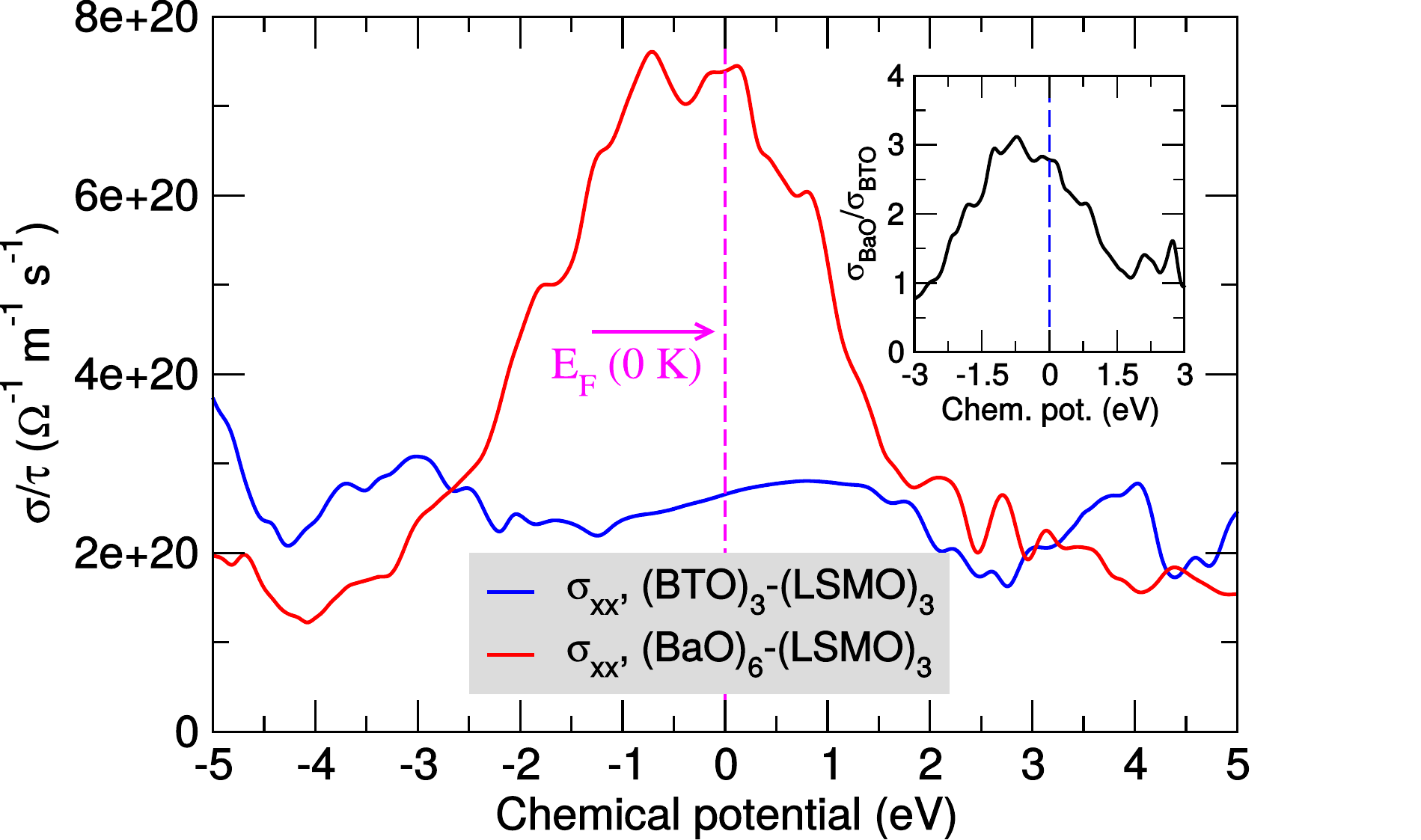}}  
\caption{Transport calculations on $\rm [LSMO]_3[BaO]_6$ and $\rm
  [LSMO]_3[BaTiO_3]_3$ systems. The dotted line represent the Fermi level.}
  \label{fig:transp}
\end{figure}

The fact that the FM and A-type AFM orders are found so close in
energy tells us that, in real systems such superlattices may present
one or the other spin arrangement as the ground state, according to
the specific cation disorder. In any way at room temperature both
state can be expected to be occupied with similar probabilities.  Our
results thus show that such superlattices should display a net total
magnetization (even-though reduced compared to the FM state), and more
importantly a large magnetic moment for the interface layers. 
Concerning the transport properties we computed the conductivity
  tensor for the $\rm [LSMO]_3[BaO]_6$ and the $\rm
  [LSMO]_3[BaTiO_3]_3$ systems using the Boltztrap~\cite{Boltz}
  code. Figure~\ref{fig:transp} clearly shows a strong increase in
  the in-plane conductivity for the system with $\rm BaO$ alternating
  layers.
The dominant $d_{x^2-y^2}$ orbital occupancy at the interfaces,
supported by the conductivity calculations, lead us to think
that using simple oxides as alternating layers is indeed a promising
way to prevent the ``dead layer'' phenomenon.

One can however wonder if this conclusion will remain valid if one
increases the size of the manganite layer. In order to check this
point we increased the size of the LSMO layer to 6 mono-layers, and
performed the calculations for one typical cation configuration. To
keep this calculation to a reasonable size, we needed to
simultaneously decrease the BaO layer thickness. We thus first checked
whether such a reduction would affect the results. For this purpose we
run test calculations on the preceding superlattice with only two
mono-layers of BaO ({\em i.e.} on $\rm [LSMO]_3 [BaO]_{\bf 2}$).
These calculations showed a similar behavior to the calculations with
6 BaO mono-layers, and thus validate 2 BaO mono-layers model.

Our calculations on the $\rm [LSMO]_{\bf 6} [BaO]_{2}$
superlattices showed that the ground state again imposes in-plane
ferromagnetism.  The spin arrangement in the {\bf c} direction displays a
$\uparrow \uparrow \downarrow \downarrow \uparrow \uparrow$ pattern
(``{\bf uudduu}'') with a total net magnetization for the system. This
ground state is again very close in energy to the FM state and the
A-type AFM state. The latter does not however correspond to a full AFM
state, since it exhibits a non null net total magnetization of about
$1/10$ of an electron per Mn atom.

The dominant $e_g$ orbital occupancy in the different LSMO mono-layers
is found qualitatively independent of the out-of--plane spin ordering
(see table~\ref{t:62} for an example). Indeed, as in the 
$\rm [LSMO]_3[BaO]_6$
calculations, the mono-layers at the interfaces are contracted and
strongly dominated by the $d_{x^2-y^2}$ orbital occupancy. In fact
only the inner most mono-layer is still elongated and dominated by
$d_{z^2}$ orbital occupancy. As it can be seen in Table~\ref{t:62}, the Mn
magnetic moments and the amplitude of the Jahn-Teller distortion
exhibit a strong correlation. The Mn-octahedra in the inner most
mono-layer exhibit a strong elongation and the largest Mn magnetic
moment. This specificity of the inner most mono-layer is responsible
for the non-vanishing total magnetization in the A-type AFM state.
\begin{table}[h!]
  \centering
  \begin{tabular}{crr@{\hskip 5ex}rrr}\hline\hline
    LSMO  & \multicolumn{2}{c}{$e_g$ orb. spin pop.\hspace{5ex}} & $c/a-1$ & 
JTd & $\mu_{\rm Mn}$ \\
    mono-layer &$d_{x^2-y^2}$ & $d_{z^2}$ & \\
    \hline
    1 & {\bf 0.46} & 0.32 & -0.032 & 0.017 & 3.55\\
    2 & {\bf 0.47} & 0.24 & -0.028 & -0.034 & 3.43 \\
    3 &{\bf -0.46} &-0.25 & -0.021 & -0.038 & -3.45\\
    4 &-0.30 &{\bf -0.74} &  {\bf 0.013} & {\bf 0.070} & -3.89 \\
    5 &{\bf  0.48} & 0.22 & -0.027 & -0.035 & 3.44\\
    6 &{\bf  0.50} & 0.29 & -0.043 & -0.012 & 3.57 \\
    \hline \hline
  \end{tabular}
  \caption{M\"ulliken spin population of the Mn $e_g$ orbitals, 
    $c/a$ ratio, JTd (Jahn-Teller distortion) and $\mu_{\rm Mn}$ (Mn magnetic 
moment) in  the  $\rm [La_{2/3}Sr_{1/3}MnO_3]_6[BaO]_2$
    ground state of one typical cation order. Values for the two other low 
energy states 
    (FM and A-type AFM) are qualitatively equivalent.
}
  \label{t:62}
\end{table}
These results show that, when increasing the thickness of the LSMO
layer, one essentially increases the thickness of the interface layers
and not of the central one. The former being contracted along {\bf c}
and dominated by $d_{x^2-y^2}$ orbital occupancy, it confirms that the
use of BaO alternating layers hinder the formation of a ``dead layer'' at
the LSMO interfaces.

Finally we checked whether this result is resilient to a change in the
simple oxide and manganite compounds. We thus performed a set of
calculations using BaO, SrO and MgO as alternating layers, and LSMO or
LBMO as manganite layers ($\rm [La_{2/3}A_{1/3}MnO_3]_3[BO]_6$), for a
typical cation disorder model~\cite{note}. Table~\ref{t:autres} summarizes the $e_g$ orbital occupancies for those calculations.
\begin{table}[h!]
  \centering
  \begin{tabular}{ccrr}\hline\hline
   & LAMO  & \multicolumn{2}{c}{$e_g$ orb. spin pop.} \\
   & mono-layer &$d_{x^2-y^2}$ & $d_{z^2}$ \\
    \hline 
    & 1 & {\bf 0.44} & 0.29 \\
   LBMO-BaO & 2 & 0.38 & {\bf 0.54}\\
   & 3 & {\bf 0.51} &   0.26 \\
   \hline
   & 1 & {\bf 0.48} & 0.27 \\
   LBMO-BaO ($P4/mmm$) & 2 & 0.24 & {\bf 0.66}\\
   & 3 & {\bf 0.48} &   0.27 \\
    \hline 
     & 1 & {\bf 0.48} & 0.26 \\
   LSMO-SrO & 2 & 0.40 & {\bf 0.52} \\
   & 3 & {\bf 0.48} & 0.26  \\
   \hline 
     & 1 & {\bf 0.48} & 0.25 \\
   LSMO-SrO ($P4/mmm$) & 2 & 0.40 & {\bf 0.53} \\
   & 3 & {\bf 0.48} & 0.25  \\
   \hline
     & 1 & {\bf 0.35} & 0.17 \\
   LSMO-MgO 
   & 2 & 0.23 & {\bf 0.70}\\
   & 3 & {\bf 0.60} & 0.26 \\
    \hline
     & 1 & {\bf 0.35 }& 0.17\\
   LBMO-MgO 
   & 2 & 0.24 &{\bf 0.70}\\
   & 3 & {\bf0.59} & 0.32 \\
    \hline\hline
  \end{tabular}
  \caption{M\"ulliken spin population of the Mn $e_g$ orbitals in  the 
    $\rm [La_{2/3}A_{1/3}MnO_3]_3[BO]_6$ ground state 
    (A=Sr,\,Ba~; B=Ba,\,Sr,\,Mg).
    The shown example was chosen as the cation ordering associated 
    with the lowest ground state energy.}
    \label{t:autres}
\end{table}
One may notice that the $\rm (LBMO)_3(BaO)_6$ and $\rm
(LSMO)_3(SrO)_6$ superlattices have in theory equivalent interfaces, unlike
all the other superlattices we studied. One sees in
table~\ref{t:autres} that this symmetry is kept in the $\rm
(LSMO)_3(SrO)_6$ superlattice. Indeed, the two calculations with and
without imposed symmetry yield equivalent results within error bars. 
For the $\rm (LBMO)_3(BaO)_6$ superlattice however, this is
not the case. Indeed, a spontaneous symmetry breaking occurs along
the {\bf c} axis, associated with a small energetic stabilization
($37\,\rm meV \simeq 430\,K$) per LBMO u.c. This induces a
symmetry breaking in the $e_g$ orbitals occupancies as can be seen in
table~\ref{t:autres}.
Nevertheless,  all manganite interface mono-layers are
favoring a $d_{x^2-y^2}$ occupancy over a $d_{z^2}$ one, as was the
case for the $\rm (LSMO)_n(BaO)_p$ compounds.  This result thus seems
to remain valid independently of the manganite compound and of the
simple oxide chosen for the alternating layer. 

As a conclusion one may recall that thin films and superlattices of
$\rm [La_{2/3}A_{1/3}MnO_3]$ (A=Sr,\,Ca) manganite compounds, over an
$\rm SrTiO_3$ substrate, have been extensively studied in the hope to
find a good material for electronic and spintronics
applications. Indeed, on such an STO, the LSMO is under
tensile strain, so one is entitled to expect that the elastic energy will favor a 
contraction
of the mono-layers along the {\bf c} direction. Due to the degeneracy
of the $e_g$ orbitals, such a contraction would have enhanced the
occupation of the $d_{x^2-y^2}$ over the $d_{z^2}$ and thus the
ferromagnetic and metallic behavior through the double exchange
mechanism. Unfortunately the formation of a non-magnetic and
insulating layer (called ``dead layer'') at the interface prevents to
reach this goal.  This ``dead layer'' originates in a weak delocalization
of the Mn $d_{z^2}$ orbitals in the empty Ti ones. The energy gain in
this phenomenon overvalues the elastic energy loss~\cite{LSMO}. As a consequence a
preferred occupancy of the Mn $d_{z^2}$ orbitals associated with an
elongation (along the {\bf c} direction) of the interface mono-layers
takes place. 

In this paper, we theoretically studied different possibilities to
hinder the interface delocalization using suitable alternating layers
in superlattices.  Our first principle calculations show that
superlattices alternating manganite and alkaline-earth simple oxides
efficiently prevent inter-layer delocalization, promote mono-layers
contraction at the interfaces and a preferred $d_{x^2-y^2}$ occupancy
over the $d_{z^2}$ one, and finally strongly increase the
  in-plane conductivity. Our studies show that this result should
hold for different manganite and alternating layer thicknesses. One
can thus reasonably expect that such superlattices may
present the long searched magnetic and electric properties.

\section*{Acknowledgments}
We thank the IDRIS (project n$^\circ$1842) and CRIHAN (project
n$^\circ$013) French supercomputer centers, the PRACE projects 
Theo-MoMuLaM and TheDeNoMo, the C\'eci facilities funded by 
F.R.S-FNRS (Grant No 2.5020.1) and Tier-1 supercomputer of the 
F\'ed\'eration Wallonie-Bruxelles funded by the Waloon Region 
(Grant No 1117545) for providing us with computer hours.
A.~B. Ko\c cak thanks the Erasmus-Mundus program of the EEC and 
the IDS-FunMat consortium for her PhD funding under project n$^\circ$2012-11.



\begin{thebibliography}{9999999}
\bibitem{note} Although the lattice mismatch of the LSMO 
and LBMO with MgO (and even SrO) might be far too large for such superlattices 
to be realized in practice, this constitues a valuable computer experiment to 
illustrate the generality of our results.
\end{thebibliography}
\end{document}